\documentstyle[prl,aps,twocolumn,amsfonts,amssymb]{revtex}
\begin{document}

\pagestyle{myheadings}
\markright{Physica Scripta \textbf{55}, 129 (1997). }
\title{A comparison of two discrete mKdV equations}

\author{C. Chandre}
\address{Laboratoire de Physique, CNRS, Universit\'e de Bourgogne,
BP 400, F-21011 Dijon, France}
\address{e-mail: cchandre@jupiter.u-bourgogne.fr}
\maketitle
\begin{abstract}
{\em Abstract}--- We consider here two discrete versions of the modified KdV equation. In one
case, some solitary wave solutions, B\"acklund transformations and integrals 
of motion are known. In the other one, only solitary wave solutions were given,
and we supply the corresponding results for this equation. We also derive the 
integrability of the second equation and give a transformation which links the 
two models.
\end{abstract}

\vspace*{.2cm}

{\em Introduction}--- In a recent paper, Takeno~\cite{take} considered two nonlinear excitation 
transfer models in a $d$-dimensional version of the simple cubic lattice 
with nearest neighbour interactions. These models highlight two discrete 
versions of the mKdV equation
\begin{eqnarray}
\dot{\phi_n}=(\phi_{n+1}-\phi_{n-1})(1+ \phi_n ^2) ,\\
\dot{\phi_n}=(\phi_{n+1}-\phi_{n-1})(1+\phi_n )^2 .
\end{eqnarray}
The first equation which is integrable has been widely studied 
(see \cite{xiao,yang,ablo}) and is known to be a discrete mKdV equation. 
But, the second equation is also related to the continuous mKdV equation 
(see \cite{take,homm}). Indeed, in the continuous limit (the lattice spacing 
$h$ becomes zero and $x=nh$), $(1)$ and $(2)$ become
\begin{eqnarray}
&& \frac{\partial \Phi}{\partial t}=2(1+\Phi^2 )\frac{\partial \Phi}{\partial x}
+\frac{1}{3}\frac{\partial ^3 \Phi}{\partial x^3} ,\\
&&\frac{\partial \Phi}{\partial t}=2(1+\Phi )^2 \frac{\partial \Phi}{\partial x}
+\frac{1}{3}\frac{\partial ^3 \Phi}{\partial x^3} ,
\end{eqnarray}
respectively. Applying the transformation $\Phi(x,t)=\Psi(\xi,t)-1$ where 
$\xi=x-2t$, $(4)$ becomes $(3)$. Therefore, the continuous versions of $(1)$ 
and $(2)$ are linked by a simple Galilean transformation. But, this 
transformation can not be applied in the discrete case. One result of this 
paper is to find the transformation which links $(1)$ and $(2)$.\\
\\
{\em Travelling wave solutions}--- For completeness, we first review 
the known travelling wave solutions of $(1)$ and $(2)$ of the form 
$ \phi_n(t)=\Phi(n-c t )$. Some solutions are given by 
Takeno~\cite{take,homm} for both equations and Xiao~\cite{xiao} for eq.$(2)$.
Both equations have bright solitary wave solutions with a localised shape; 
for $(1)$ and $(2)$ respectively we have~\cite{take}
\begin{eqnarray}
   &&\phi_n=\pm \,\sinh (C) \, \mathrm{sech} [\it C(n-n_0)-\omega t],\\ 
   && \omega=-2\sinh C , \\
   &&\phi_n= \,\sinh ^2 (C) \, \mathrm{sech} ^2 [\it C(n-n_0)-\omega t],\\
   && \omega=-\sinh 2C .
\end{eqnarray}
Besides, $(2)$ has the singular solution~\cite{take}
$$ \phi_n= \, -\sinh ^2 (C) \, \mathrm{cosech} ^2 [\it C(n-n_0)-\omega t] .$$
We notice that the solitary wave solutions of $(1)$ are closer to the
continuous case which has sech-shaped solitary wave solutions. But, $(2)$ is
interesting because it has non vanishing waves with a sech$^2$ shape. 
Xiao~\cite{xiao} also gives solitary wave solutions with nonzero boundary 
conditions of $(2)$ by using the real exponential approach
\begin{eqnarray*}
&&\phi_n= a_0 + (1+a_0) \,\sinh ^2 (C) \, \mathrm{sech} ^2 
[\it C(n-n_0)-\omega t], \\
&&\phi_n= a_0 - (1+a_0) \,\sinh ^2 (C) \, \mathrm{cosech} ^2 
[\it C(n-n_0)-\omega t] , 
\end{eqnarray*}
where $\omega=-(1+a_0) ^2 \sinh 2C$.
For $(1)$, there is no known corresponding solutions with a sech shape with 
nonzero boundary conditions.\\

{\em Lax pairs}---
We now consider the Lax pairs for $(1)$ and $(2)$. Recall that for discrete 
equation $ \partial \phi_n/\partial t=F(\phi_n,\phi_{n-1},\phi_{n+1})$,
a Lax pair is a set of two matrices $( M_n , N_n )$ which satisfy
\begin{eqnarray*}
         && V_{n+1} = M_n V_n, \\
         && \partial V_n/\partial t = N_n V_n,
\end{eqnarray*}
where $ N_n =\displaystyle \left( \begin{array}{cc} 
    A_n & B_n \\ C_n & D_n \end{array} \right)$ and $ V_n=
    \displaystyle\left( 
    \begin{array}{ll} V_{n,1} \\ V_{n,2} \end{array} \right) ,$
 and also, the compatibility condition 
 $$ \frac{\partial}{\partial t} \left( S V_n \right) = S 
 \frac{\partial V_n}{\partial t},$$ 
 where $S$ is  the shift operator $ S V_n = V_{n+1}$.
In this case, this condition is $ 
\partial M_n/\partial t= N_{n+1} M_n - M_n N_n $ and gives the 
dynamical equation for $\phi_n$.
For Equation $(1)$, the Lax pair is given by Ablowitz and Segur~\cite{ablo2}:
$$M_n =\left( \begin{array}{cc} \eta & \phi_n \\ -\phi_n & 1/\eta 
\end{array} \right) \indent N_n =\left( \begin{array}{cc} A_n & B_n \\ 
C_n & D_n \end{array} \right),$$
\begin{eqnarray*}
 \mbox{ where } &&  A_n= \eta^2 +\phi_{n-1} \phi_n, \indent 
                    B_n=\phi_n \eta + \phi_{n-1}/\eta, \\
 && C_n=-\phi_{n-1} \eta - \phi_n/\eta, \indent  
   D_n= \eta^{-2} +\phi_{n-1} \phi_n,
\end{eqnarray*}
and $\eta$ is the spectral parameter associated with the problem.
It is straightforward to show that the Lax pair for $(2)$ is
$$ M_n =\left( \begin{array}{cc} \eta & 1+\phi_n \\ -(1+\phi_n) & 0 
\end{array} \right), \indent N_n =\left( \begin{array}{cc} A_n & B_n \\ 
C_n & D_n \end{array} \right), $$
\begin{eqnarray*}
\mbox{ where } && A_n=(1+\phi_n) (1+\phi_{n-1}) +\eta^2, \indent 
          B_n=\eta (1+\phi_n). \\ 
       && C_n=-\eta(1+\phi_{n-1} ), \indent D_n= (1+\phi_n ) (1+\phi_{n-1}).
\end{eqnarray*}
This proves the complete integrability of both differential-difference 
equations.

{\em Integrals of motion}---
 We now consider integrals of motion, which are easy to obtain 
 (see Ablowitz and Ladik~\cite{ablo}, Homma and Takeno~\cite{homm}, Scharf 
 and Bishop~\cite{scha}) for an infinite 1D-lattice for $(1)$
 \begin{equation}  N_1=\sum_{n=-\infty}^{+\infty} 
 \ln (1+\phi_n ^2) ,\indent   N_1 '=\sum_{n=-\infty}^{+\infty}
 \arctan \phi_n .\end{equation}
In a similar way for eq.$(2)$, we have
 \begin{equation} N_2=\displaystyle \sum_{n=-\infty}^{+\infty} 
 \ln (1+\phi_n ) ,\indent N_2 '=\displaystyle \sum_{n=-\infty}^{+\infty}
 \frac{\phi_n}{1+\phi_n} .\end{equation} 
Other conserved quantities of $(1)$ are given by Ablowitz and 
Ladik~\cite{ablo}: for instance,
\begin{eqnarray} 
&& C_1 =  \sum_{n=-\infty}^{+\infty} \phi_n \phi_{n-1} ,\\
&& C_2 =  \sum_{n=-\infty}^{+\infty} \left\{ \phi_n \phi_{n-2} 
( 1+\phi_{n-1}^2) + \phi_n^2 \phi_{n-1}^2/2 \right\}.
\end{eqnarray}

{\em B\"acklund transformations}---
The B\"acklund transformation for $(1)$ was derived by 
Yang and Schmid~\cite{yang}, who give the two following Miura transformations
\begin{eqnarray}
   && \phi_n=\frac{u_{n+1}/\eta -u_n \eta}{1+u_n u_{n+1}} 
   = \Theta_1 (u_n,u_{n+1}) ,\\ 
   && \phi_{n+1} '=\displaystyle \frac{u_{n}/\eta 
-u_{n+1} \eta}{1+u_n u_{n+1}} = \Theta_2 (u_n,u_{n+1}).
\end{eqnarray} 
where $u_n=V_{n,1}/V_{n,2}$ is a solution of the following 
differential-difference equation
\begin{eqnarray*}
\frac{\partial u_n}{\partial t} =&& \displaystyle \frac{u_{n+1} -u_n 
\left( \eta^2 +\eta^{-2} \right) -u_n ^3}{1+u_n u_{n+1}}\\
&& - \frac{u_{n-1} -u_n \left( \eta^2 + \eta^{-2} \right)
-u_n ^3}{1+u_n u_{n-1}}
\end{eqnarray*}
Therefore, $ \Phi ' = (\Theta_2 \circ \Theta_1 ^{-1} ) \Phi $ defines a  
B\"acklund transformation for $(1)$.

Applying the same method to $(2)$, we find two Miura transformations which 
define a B\"acklund transformation for this equation 
\begin{eqnarray} 
&& \phi_n= \displaystyle -1 -\eta 
\frac{v_n}{1+v_n v_{n+1}}=\tilde{\Theta_1} (v_n,v_{n+1}) ,\\ 
&& \phi_{n+1} '=\displaystyle -1 -\eta \frac{v_{n+1}}{1+v_n v_{n+1}} =
\tilde{\Theta_2} (v_n,v_{n+1}).
\end{eqnarray}
where $v_n=V_{n,1}/V_{n,2}$ is a solution of the following 
differential-difference equation:
$$
 \frac{\partial v_n}{\partial t}= \eta ^2 v_n ^2 
 \left( \frac{v_{n+1}}{1+v_n v_{n+1}} -\frac{v_{n-1}}{1+v_n v_{n-1}} \right) 
$$
Therefore, $ \Phi ' = (\tilde{\Theta_2} \circ \tilde{\Theta_1} ^{-1} ) \Phi $ 
defines a  B\"acklund transformation for $(2)$.\\
These B\"acklund transformations link two solutions of the same 
differential-difference equation.

{\em Miura-type transformation}---
Finally, we derive a transformation which links the two discrete mKdV 
equations under consideration, in the same way as their continuous versions 
are linked by a Galilean transformation. Unfortunately, some features of $(2)$ 
are not similar to those of $(1)$ (for instance, the travelling wave solutions 
$(5,6)$  or the integrals of motion $(7,8)$). So, the transformation which 
links both discrete equations is more difficult to obtain than the continuous 
case.\\
The idea is to link both equations with a discrete version of the KdV 
equation by two Miura~transformations. \\
If we assume that 
\begin{equation} u_n = (1+i \phi_n) (1-i \phi_{n+1}), \end{equation}  $(1)$ 
becomes by a straightforward calculation
\begin{equation}
\dot{u_n} = u_n (u_{n+1} -u_{n-1} ) .
\end{equation}
This defines a Miura transformation $\hat{\Theta_1}$ which maps a solution of 
$(13)$ into a solution of $(1)$. \\
On the other hand, if we define 
\begin{equation} u_n = (1+ \phi_n) (1+ \phi_{n+1}) , \end{equation} 
$(2)$ becomes $(13)$. This also defines a Miura transformation  
$\hat{\Theta_2}$ which maps a solution of  $(13)$ into a solution of $(2)$. 
Therefore, the transformation 
$\Phi ' = (\hat{\Theta_2} \circ \hat{\Theta_1} ^{-1} ) \Phi $ 
maps a solution of $(1)$ into a solution of $(2)$. In theory, this 
transformation could be used to derive the results above, but such 
calculations do not appear to be straightforward.

{\em Acknowledgements}--- I am grateful to J.C. Eilbeck 
for suggesting this problem to me, and for the hospitality of the 
Department of Mathematics, Heriot-Watt University, where much of this 
work was carried out. The visit was supported by funds from the ENS.


\begin{references}

\bibitem{take}
S.~Takeno,
\newblock { Phys. Soc. Japan} {\bf 61}, 1433 (1992).

\bibitem{xiao}
Y.~Xiao.
\newblock { Phys. Lett. A} {\bf 193}, 419 (1994).

\bibitem{yang}
X.~Yang and R.~Schmid.
\newblock { Phys. Lett. A} {\bf 195}, 63 (1994).

\bibitem{ablo}
M.J. Ablowitz and J.F. Ladik.
\newblock { J. Math. Phys} {\bf 17}, 1011 (1976).

\bibitem{homm}
S.~Homma and S.~Takeno.
\newblock { Phys. Lett. A} {\bf 169}, 355 (1992).

\bibitem{ablo2}
M.J. Ablowitz and H.~Segur.
\newblock { Solitons and the inverse scattering transform}.
\newblock (SIAM, Philadelphia, 1981).

\bibitem{scha}
R.~Scharf and A.R. Bishop.
\newblock { Phys. Rev. A} {\bf 43}, 6535 (1991).

\end{references}
\end{document}